\documentstyle[sprocl]{article}

\def\Journal#1#2#3#4{{#1} {\bf #2}, #3 (#4)}

\def\NPB{{\it Nucl. Phys.} B}

\def\PRL{\it Phys. Rev. Lett.}
\def\PRD{{\it Phys. Rev.} D}

\def\be{\begin{equation}}
\def\ee{\end{equation}}
\def\bea{\begin{eqnarray}}
\def\eea{\end{eqnarray}}

\begin{document}

\title{TESTING CPT AND LORENTZ INVARIANCE WITH THE ANOMALOUS SPIN PRECESSION 
OF THE MUON}
\author{M.~DEILE$^{11}$, M.~BAAK$^{11}$,
H.N.~BROWN$^2$, G.~BUNCE$^2$, R.M.~CAREY$^1$,
P.~CUSHMAN$^{9}$, G.T.~DANBY$^2$, 
P.T.~DEBEVEC$^7$, H.~DENG$^{11}$, W.~DENINGER$^7$, 
S.K.~DHAWAN$^{11}$, V.P.~DRUZHININ$^3$,
L.~DUONG$^{9}$, E.~EFSTATHIADIS$^1$, F.J.M.~FARLEY$^{11}$, 
G.V.~FEDOTOVICH$^3$, S.~GIRON$^{9}$,
F.~GRAY$^7$, D.~GRIGORIEV$^3$, M.~GROSSE-PERDEKAMP$^{11}$,
A.~GROSSMANN$^6$, M.F.~HARE$^1$, D.W.~HERTZOG$^7$, 
V.W.~HUGHES$^{11}$, M.~IWASAKI$^{10}$,
K.~JUNGMANN$^6$, D.~KAWALL$^{11}$,
M.~KAWAMURA$^{10}$, B.I.~KHAZIN$^3$, J.~KINDEM$^{9}$, F.~KRIENEN$^1$,
I.~KRONKVIST$^{9}$, R.~LARSEN$^2$,
Y.Y.~LEE$^2$, I.~LOGASHENKO$^{1,3}$, R.~MCNABB$^{9}$, W.~MENG$^2$, 
J.~MI$^2$, J.P.~MILLER$^1$, W.M.~MORSE$^2$,
D.~NIKAS$^2$, C.J.G.~ONDERWATER$^7$, Y.~ORLOV$^4$, C.S.~\"{O}ZBEN$^2$,
J.M.~PALEY$^1$, C.~POLLY$^7$,
J.~PRETZ$^{11}$, R.~PRIGL$^2$, G.~ZU~PUTLITZ$^6$, S.I.~REDIN$^{11}$,
O.~RIND$^1$, B.L.~ROBERTS$^1$, N.~RYSKULOV$^3$,
S.~SEDYKH$^7$, Y.K.~SEMERTZIDIS$^2$, YU.M.~SHATUNOV$^3$, 
E.P.~SICHTERMANN$^{11}$, E.~SOLODOV$^3$,
M.~SOSSONG$^7$, A.~STEINMETZ$^{11}$, L.R.~SULAK$^1$,
C.~TIMMERMANS$^{9}$, A.~TROFIMOV$^1$,
D.~URNER$^7$, P.~VON~WALTER$^6$, D.~WARBURTON$^2$, D.~WINN$^5$, 
A.~YAMAMOTO$^8$, D.~ZIMMERMAN$^{9}$}
\address{
$^1$Department of Physics, Boston University, Boston, MA 02215, USA\\
$~^2$Brookhaven National Laboratory, Upton, NY 11973, USA\\
$~^3$Budker Institute of Nuclear Physics, Novosibirsk, Russia\\
$~^4$Newman Laboratory, Cornell University, Ithaca, NY 14853, USA\\
$~^5$Fairfield University, Fairfield, CT 06430, USA\\
$~^6$Physikalisches Institut der Universit\"{a}t Heidelberg, 69120
Heidelberg, Germany\\
$~^7$Department\,of\,Physics,\,University\,of\,Illinois\,at\,Urbana-Champaign,\,IL\,61801,\,USA\\
$~^8$KEK, High Energy Accelerator Research Organization, \\Tsukuba, Ibaraki
305-0801, Japan\\
$~^{9}$Department of Physics, University of Minnesota, Minneapolis, MN
55455, USA\\
$~^{10}$Tokyo Institute of Technology, Tokyo, Japan\\
$~^{11}$Department of Physics, Yale University, New Haven, CT 06520, USA
}


\maketitle\abstracts{
This article discusses tests of CPT and Lorentz invariance with data from the
muon g-2 experiment at Brookhaven National Laboratory.
According to an extension of the Standard
Model by Kosteleck\'{y} et al., CPT/Lorentz
violating terms in the Lagrangian induce a shift of the anomaly frequency
$\omega_a$ of muons in a magnetic field. This
shift is predicted to be different for positive and negative muons
and to oscillate with the Earth's sidereal frequency. 
We discuss the sensitivity of the g-2 experiment to different parameters of 
this Standard Model extension and propose an analysis method to search 
for sidereal variations of $\omega_a$. 
}

%
%
\section{The Muon g-2 Experiment at BNL}
The muon g-2 experiment 
determines the $g$ value anomaly $a_{\mu} = (g - 2) / 2$ of the muon 
by measuring the difference $\omega_a$ between the spin precession angular
frequency $\omega_s$ and the cyclotron angular frequency $\omega_c$
of highly polarized muons of momentum 3.09\,GeV/$c$ in a 
14.2\,m diameter storage ring with a 
homogeneous magnetic field of 1.45\,T strength. The field is determined
from the NMR frequency
of protons in water, calibrated relative to the free proton NMR frequency
$\omega_{p}$.
The anomaly $a_{\mu}$ is determined via
\begin{equation}
a_{\mu} = \frac{\omega_a / \omega_p}{\mu_\mu / \mu_p
- \omega_a / \omega_p}
\end{equation}
where the ratio $\mu_\mu / \mu_p$ is taken from the measurement
by W. Liu et al.~\cite{lambda}
For muons with $\gamma = 29.3$, $\omega_a$ is not affected by the electrostatic
field from the quadrupoles used for vertical focussing.
Experimental details and the result for
$a_{\mu^{+}}$ based on the data of 1999 with an uncertainty of
1.3\,ppm are described elsewhere.~\cite{prl}
%
%
\section{Predicted Effects of CPT/Lorentz Violation on $\mathbf{\omega_a}$}
Kosteleck\'{y} et al.~\cite{smextension} suggested an extension of the 
Standard Model which
retains SU(3)$\times$SU(2)$\times$U(1) gauge invariance,
renormalizability and energy conservation, but allows for violation of CPT and 
Lorentz invariance. 
The additional muon terms in the Lagrangian 
are given by~\cite{kostbluhm}
\begin{equation}
\label{eqn_lagrange}
\begin{array}{rcl}
{\mathcal L'} &=& - a_{\kappa} \bar{\psi} \gamma^{\kappa} \psi
- b_{\kappa} \,\bar{\psi} \gamma_{5} 
\gamma^{\kappa} \psi 
 - \frac{1}{2} H_{\kappa \lambda} \,
\bar{\psi} \sigma^{\kappa \lambda} \psi \\
&&+ \frac{1}{2} i c_{\kappa \lambda} \bar{\psi} \gamma^{\kappa} 
\stackrel{\leftrightarrow}{D^{\lambda}} \psi
+ \frac{1}{2} i \, d_{\kappa \lambda} \,
\bar{\psi} \gamma_{5} \gamma^{\kappa} 
\stackrel{\leftrightarrow}{D^{\lambda}} \psi \quad .
\end{array}
\end{equation}
All terms violate Lorentz symmetry whereas CPT is only broken by the terms 
involving $a_{\kappa}$
and~$b_{\kappa}$. The additional terms are expected to be very small, based
on established experimental bounds on CPT violation for other particles like
$K^{0}$ -- $\bar{K}^{0}$~\cite{pdg} or the electron.~\cite{dehmelt} 
From Eq.~(\ref{eqn_lagrange}) the correction to $\omega_{a}$ 
can be calculated.~\cite{kostbluhm} The result is
\begin{equation}
\label{eqn_shifts_cel}
\delta \omega_{a}^{\mu^{\pm}} = 2 \check{b}_{Z}^{\mu^{\pm}} \cos \chi + 
2 (\check{b}_{X}^{\mu^{\pm}} \cos \Omega t + 
   \check{b}_{Y}^{\mu^{\pm}} \sin \Omega t) \sin \chi
\end{equation}
where 
\begin{equation}
\label{eqn_defcheck}
\check{b}_{J}^{\mu^{\pm}} \equiv \pm \frac{b_{J}}{\gamma} + m_{\mu} d_{J0} 
+ \frac{1}{2} \varepsilon_{JKL} H_{KL} \quad .
\end{equation}
Here, the model parameters are expressed in terms of a non-rotating celestial 
frame of reference whose $Z$-direction is oriented along the Earth's 
rotational axis. The angle $\chi$ is the geographic
colatitude at the experiment, i.e. the angle between 
the vertical direction of the laboratory frame and the Earth's axis.
Due to the rotation of the Earth, $\delta \omega_{a}^{\mu^{\pm}}$
has a component oscillating with the sidereal frequency
$\Omega \equiv 2 \pi / 23{\rm h} 56{\rm min}$.
Eq.~(\ref{eqn_shifts_cel}) predicts two experimental signatures of CPT/Lorentz 
violation, discussed below.
%
%
\subsection{Sidereal time dependence of $\omega_{a}^{\mu^{+}}$ and
$\omega_{a}^{\mu^{-}}$}
A g-2 experiment measuring $\omega_{a}$ for $\mu^{+}$ or $\mu^{-}$ over
a period of several days is sensitive to the oscillating terms in 
Eq.~(\ref{eqn_shifts_cel}).
From the oscillation amplitude $\hat{\omega}_{a}^{\mu^{+}}$ or   
$\hat{\omega}_{a}^{\mu^{-}}$
a combination of model parameters can be extracted:
\begin{equation}
\sqrt{(\check{b}_{X}^{\mu^{\pm}})^{2} + (\check{b}_{Y}^{\mu^{\pm}})^{2}}
= \frac{\hat{\omega}_{a}^{\mu^{\pm}}}{2 \, |\sin \chi|}
\end{equation}
Since $\omega_{a}$ is proportional to the magnetic field $B$, the test has to
be performed at constant $B$. 

The parameter combination probed with this test is very similar to the one
determined from muonium spectroscopy.~\cite{muonium} The only difference 
is that for muonium the quantities $\check{b}_{J}^{\mu^{\pm}}$,
defined in Eq.~(\ref{eqn_defcheck}), are replaced by their non-relativistic
limits with $\gamma = 1$.~\cite{kostbluhm} 
Therefore the sensitivity of g-2 tests to the basic parameter 
components $b_{X}$ and $b_{Y}$ is suppressed by the factor $\gamma=29.3$ 
with respect to muonium, for a given sensitivity to
$\check{b}_{X}^{\mu^{\pm}}$ and $\check{b}_{Y}^{\mu^{\pm}}$.

An analysis procedure for sidereal variations is currently being developed 
using the $\mu^{+}$ data of 1999 whose
uncertainty for the time average of $\omega_{a}$ is 
$\sigma(\langle\omega_{a}\rangle) = 1.3\,$ppm. The error on
the sidereal oscillation amplitude $\hat{\omega}$ is determined analytically
from the 
second derivatives of $\chi^{2}$ for the fit function~(\ref{eqn_fit}) discussed
in Section~\ref{sec_analysis}. A simple calculation yields
$\sigma(\hat{\omega}_{a}) \approx \sqrt{2} \sigma(\langle\omega_{a}\rangle) =
1.8$\,ppm for the 1999 data. This leads to an achievable uncertainty for
$\sqrt{(\check{b}_{X}^{\mu^{\pm}})^{2} + (\check{b}_{Y}^{\mu^{\pm}})^{2}}$
of about 1.2$\times 10^{-24}$ GeV. Taking into account the above-mentioned 
suppression by $\gamma=29.3$, the parameters
$b_{X}$ and $b_{Y}$ can be probed with a sensitivity of the order 
3.5$\times 10^{-23}$\,GeV, whereas muonium tests achieve a limit of 
2$\times 10^{-23}$\,GeV.

To quantify the level of CPT/Lorentz violating effects, the dimensionless 
figure of merit $r_{\hat{\omega}_{a}} \equiv \frac{\hat{\omega}_{a}}{m_{\mu}}$
was introduced~\cite{kostbluhm}, interpreting
$\delta \omega_{a}$ as an energy shift of the muon and comparing its amplitude 
with the rest energy $m_{\mu}$. With the 1999 data set
$r_{\hat{\omega}_{a}}$ can be probed down to a level of 0.19$\times 10^{-22}
\ll \frac{m_{\mu}}{m_{\rm Planck}}$, implying sensitivity to physics 
beyond the Planck scale. With the total data set for $\mu^{+}$ the sensitivity
will be significantly better since in 2000 approximately four times more data 
were recorded. The final statistics for $\mu^{-}$ are expected to be similar.

\subsection{Comparison of $\omega_{a}$ for positive and negative muons}
From Eq.~(\ref{eqn_shifts_cel}) follows that two experiments at the 
same colatitude $\chi$ and at the same magnetic field will 
observe the time-averaged $\omega_{a}$-difference 
\begin{equation}
\label{eqn_mu+mu-}
\Delta \omega_{a} \equiv
\langle\omega_{a}^{\mu^{+}}\rangle - \langle\omega_{a}^{\mu^{-}}\rangle \nonumber = 
4 \frac{b_{Z}}{\gamma} \cos \chi \quad.
\end{equation}
Thus the measurement of $\Delta \omega_{a}$ sets a bound on $b_{Z}$
and is therefore complementary to the measurement of sidereal variations 
which involve the parameters $b_{X}$ and $b_{Y}$. 
The CERN g-2 experiment~\cite{cern} which measured both 
$\omega_{a}^{\mu^{+}}$ and $\omega_{a}^{\mu^{-}}$ with an uncertainty of 
10\,ppm, obtained $\Delta \omega_{a} = 2\pi (-5.5 \pm 3.3)\,$Hz, 
corresponding to $b_{Z} = (-2.3 \pm 1.4)\times10^{-22}$\,GeV, or a figure
of merit~\cite{kostbluhm}
$r_{\Delta \omega_{a}} \equiv \frac{\Delta \omega_{a}}{m_{\mu}} = 
(-2.2 \pm 1.3) \times 10^{-22}$. 
With the final statistics of the g-2 experiment at BNL the sensitivity is 
expected to improve by roughly a factor 15.

One can also compare $\omega_{a}^{\mu^{+}}$ from BNL with
$\omega_{a}^{\mu^{-}}$ from CERN. However, one needs to consider 
that the two experiments are
situated at different geo\-graphic colatitudes $\chi_{1}$ and $\chi_{2}$ 
and that the magnetic fields $B_{1}$ and $B_{2}$
were slightly different. Using Eq.~(\ref{eqn_shifts_cel}), 
the measured frequencies can be written as 
\begin{eqnarray}
\langle\omega_{a}^{\mu^{+}}\rangle = \frac{e}{m} a B_{1} + 
\langle\delta\omega_{a}^{\mu^{+}}\rangle \\
\langle\omega_{a}^{\mu^{-}}\rangle = \frac{e}{m} a B_{2} + 
\langle\delta\omega_{a}^{\mu^{-}}\rangle
\end{eqnarray}
To cancel the B-field dependent terms, we scale the CERN result
$\langle\omega_{a}^{\mu^{-}}\rangle$ by $\frac{B_{1}}{B_{2}}$ and subtract it
from $\langle\omega_{a}^{\mu^{+}}\rangle$:
\begin{eqnarray}
\label{eqn_mu+mu-cernbnl}
&&\Delta \omega_{a} =
\langle\omega_{a}^{\mu^{+}}\rangle - \frac{B_{1}}{B_{2}} 
\langle\omega_{a}^{\mu^{-}}\rangle 
\nonumber \\
&&= 2 \frac{b_{Z}}{\gamma} \left(\cos \chi_{1} + \frac{B_{1}}{B_{2}} 
\cos \chi_{2}\right) +
2 \left(m_{\mu} d_{Z0} + H_{XY}\right) \left(\cos \chi_{1} - \frac{B_{1}}{B_{2}}
\cos \chi_{2}\right)~~.~
\end{eqnarray}
The usefulness of such a test is its sensitivity to a
parameter combination involving not only $b_{Z}$ but also $d_{Z0}$ and 
$H_{XY}$:
From the 1999 result~\cite{prl} for $\omega_{a}^{\mu^{+}}$ and the 
CERN value~\cite{cern} for $\omega_{a}^{\mu^{-}}$ we obtain
\( 
b_{Z} + \gamma \frac{\cos \chi_{1} - \frac{B_{1}}{B_{2}}\cos \chi_{2}}
{\cos \chi_{1} + \frac{B_{1}}{B_{2}} \cos \chi_{2}} (m\, d_{z0} + H_{xy})
= (-1.4 \pm 1.0)\times10^{-22}\,{\rm GeV,}
\)
where $\chi_{1}$(BNL) = 49.1$^{\rm o}$, $\chi_{2}$(CERN) = 43.8$^{\rm o}$
and $\frac{B_{1}}{B_{2}} = 0.984$. This corresponds to 
$r_{\Delta \omega_{a}} = (-1.3 \pm 0.9)\times10^{-22}$.

%
\section{Proposed Analysis Procedure for Sidereal Variations of 
$\mathbf{\omega_a}$}
\label{sec_analysis}
At present a search for sidereal variations of $\omega_{a}$ in the data set
of 1999 for positive muons is underway. These data with a statistical
uncertainty of 1.25\,ppm in $\omega_{a}$ were collected in 806 runs 
distributed over 25 days. The typical length of a run was half an hour. 
While in the g-2 analysis~\cite{prl} the decay
positron time spectra of all runs were summed before the fit, 
the search for time variations in $\omega_{a}$ requires analyzing
each run individually. After applying a correction for 
pulses arriving with a time separation smaller than the resolution of the
pulse-finding algorithm, 
the g-2 analysis fits the time spectra with the function
\begin{equation}
N(t) = N_0\, {\rm e}^{-\frac{t}{\tau}} [1 + A \cos(\omega_a t + \phi)] \cdot
f_{\rm CBO}(t) \cdot f_{\rm ml}(t)
\end{equation}
The data of single runs are only sensitive to the exponential decay modulated
by the g-2 oscillation. The parameters pertaining to 
perturbations due to coherent betatron oscillations, $f_{\rm CBO}(t)$, 
and muon losses, $f_{\rm ml}(t)$, are fixed
to the values found in the g-2 analysis. This is a valid approach as long as 
$f_{\rm CBO}(t)$ and $f_{\rm ml}(t)$ are not affected by sidereal variations
from CPT/Lorentz violating effects.

The values $\omega_{a\,i}$ determined for each run at a time $t_{i}$ 
are fitted with the function
\begin{equation}
\label{eqn_fit}
\omega_{a}(t_{i}, \omega_{p\,i}) = K\, \omega_{p\,i} + \hat{\omega}_{a}
\cos(\Omega t_{i} + \Phi)
\end{equation}
taking into account the magnetic field monitored with NMR probes in terms of
the free proton precession frequency $\omega_{p}$.
$K$, $\hat{\omega}$ and $\Phi$ are free parameters whereas the sidereal
frequency $\Omega$ is fixed. 

Including the magnetic field into the fit is necessary because its variations
of up to 0.5\,ppm cause variations in $\omega_{a}$ of equal relative size.
Particularly dangerous are day-night field 
variations caused by temperature changes 
which could fake a sidereal oscillation. 

Another systematic concern involves possible effects of CPT/Lorentz violation
on the time and frequency references used in the experiment. They could
exhibit their own sidereal variations and hence either mask
the variation for muons or introduce false signatures.
A sidereal change of $\omega_{p}$ -- the basis of the NMR field measurement~-- 
could counteract with the muon-related change in $\omega_{a}$.
However, atomic clock comparisons~\cite{atomic} provide
an upper bound on the shift of $\omega_{p}$ on the mHz level, well below a
ppb of the measured frequency.
Finally, the Loran-C frequency standard used for the 
timing of the positron pulse readout and for the NMR measurement 
is based on Cesium hyperfine transitions with $m_{F} = 0$, which 
are insensitive to any preferential direction in space and will therefore not 
induce any sidereal variations. 

%
\section{Conclusions}
Tests based on muon g-2 experiments are sensitive to the muon sector of the
CPT/Lorentz violating extension~(\ref{eqn_lagrange}) of the Standard Model.
The search for sidereal variations of $\omega_{a}$ and the comparison
of $\omega_{a}$ for $\mu^{+}$ and $\mu^{-}$ probe complementary parts of
the parameter space of the extension. An analysis searching 
for sidereal variations of $\omega_{a}$ is currently being done.

%
%
\section*{Acknowledgements}
This work was supported by the U.S. Department of Energy, the
U.S. National Science Foundation, the German Bundesminister f\"ur Bildung
und Forschung, the Russian Ministry of 
Science, and the US-Japan Agreement in High Energy Physics.
Mario Deile acknowledges support by the Alexander von Humboldt Foundation.

\end{document}